\documentclass[a4paper,preprintnumbers,amsmath,amssymb,twocolumn,10pt]{revtex4}

\usepackage{graphicx}
\usepackage{dcolumn}
\usepackage{bm}
\usepackage{epstopdf}

\newcount\driver
\newcount\bozza

scaled\magstep1                  
scaled\magstep1 \font\msytw=msbm10 scaled\magstep1
 
scaled\magstep1                 \font\indbf=cmbx10 scaled\magstep2

scaled \magstep2

{\count255=\time\divide\count255 by 60
\xdef\hourmin{\number\count255}
        \multiply\count255 by-60\advance\count255 by\time
   \xdef\hourmin{\hourmin:\ifnum\count255<10 0\fi\the\count255}}

\let\a=\alpha \let\b=\beta    \let\g=\gamma     \let\d=\delta     \let\e=\varepsilon
\let\z=\zeta  \let\h=\eta     \let\th=\vartheta \let\k=\kappa     \let\l=\lambda
\let\m=\mu    \let\n=\nu      \let\x=\xi        \let\p=\pi        \let\r=\rho
\let\s=\sigma \let\t=\tau            
\let\ps=\psi   \let\o=\omega     
\let\G=\Gamma \let\D=\Delta       \let\L=\Lambda

\def\VV{{\cal V}}

\def\ZZ{{\cal Z}}
\def\RR{{\cal R}}\def\LL{{\cal L}}

\def\pp{{\bf p}}\def\qq{{\bf q}}\def\xx{{\bf x}}
  
\def\yy{{\bf y}}\def\kk{{\bf k}}\def\nn{{\bf n}}

       \def\oo{{\underline \omega}}
\def\ee{{\underline \varepsilon}}

        \def\ZZZ{\hbox{\msytw Z}}

\let\==\equiv

\let\io=\infty
\let\0=\noindent

\def\*{{\hfill\break\null\hfill\break}}

\def\media#1{{\langle#1\rangle}}

\def\eg{\hbox{\it e.g.\ }}

\def\tilde#1{{\widetilde #1}}

\def\tende#1{\,\vtop{\ialign{##\crcr\rightarrowfill\crcr
             \noalign{\kern-1pt\nointerlineskip}
             \hskip3.pt${\scriptstyle #1}$\hskip3.pt\crcr}}\,}
\def\otto{\,{\kern-1.truept\leftarrow\kern-5.truept\to\kern-1.truept}\,}

\def\Tr{\rm Tr}

\def\wh#1{\widehat{#1}}
\def\hat#1{\wh{#1}}
\def\sqt[#1]#2{\root #1\of {#2}}

\def\bp{{\bar \ps}}

\def\VV{{\cal V}}

\def\ZZ{{\cal Z}}
\def\RR{{\cal R}}\def\LL{{\cal L}}

\def\T#1{{#1_{\kern-3pt\lower7pt\hbox{$\widetilde{}$}}\kern3pt}}
\def\VVV#1{{\underline #1}_{\kern-3pt
\lower7pt\hbox{$\widetilde{}$}}\kern3pt\,}
\def\W#1{#1_{\kern-3pt\lower7.5pt\hbox{$\widetilde{}$}}\kern2pt\,}

\def\indica{\leaders \hbox to 0.5cm{\hss.\hss}\hfill}
\def\guida{\leaders\hbox to 1em{\hss.\hss}\hfill}
\mathchardef\oo= "0521

\def\V#1{{\bf #1}}
\def\pp{{\bf p}}\def\qq{{\bf q}}\def\xx{{\bf x}}
\def\yy{{\bf y}}\def\kk{{\bf k}}\def\nn{{\bf n}}

\def\oo{{\underline \omega}}

\def\qed{\raise1pt\hbox{\vrule height5pt width5pt depth0pt}}
 
  \def\bp{{\bar p}} 

\def\indic{\hbox{\raise-2pt \hbox{\indbf 1}}}

 \def\ZZZ{\hbox{\msytw Z}}

\def\ins#1#2#3{\vbox to0pt{\kern-#2 \hbox{\kern#1 #3}\vss}\nointerlineskip}

\newdimen\xshift \newdimen\xwidth \newdimen\yshift

\def\insertplot#1#2#3#4#5#6{%
\xwidth=#1pt \xshift=\hsize \advance\xshift by-\xwidth \divide\xshift by 2%
\begin{figure}[ht]
\vspace{#2pt} \hspace{\xshift}
\begin{minipage}{#1pt}
#3 \ifnum\driver=1 \griglia=#6
\ifnum\griglia=1 \openout13=griglia.ps \write13{gsave .2
setlinewidth} \write13{0 10 #1 {dup 0 moveto #2 lineto } for}
\write13{0 10 #2 {dup 0 exch moveto #1 exch lineto } for}
\write13{stroke} \write13{.5 setlinewidth} \write13{0 50 #1 {dup 0
moveto #2 lineto } for} \write13{0 50 #2 {dup 0 exch moveto #1
exch lineto } for} \write13{stroke grestore} \closeout13
\includegraphics{griglia.ps} \fi
\includegraphics{#4.ps}\fi%
\ifnum\driver=2 \fi
\end{minipage}
\caption{#5}
\end{figure}
}

\newdimen\shift \shift=-1.5truecm
\def\lb#1{%
\ifnum\bozza=1
\label{#1}\rlap{\hbox{\hskip\shift$\scriptstyle#1$}}
\else\label{#1} \fi}

\def\be{\begin{equation}}
\def\ee{\end{equation}}
\def\bea{\begin{eqnarray}}\def\eea{\end{eqnarray}}
\def\bean{\begin{eqnarray*}}\def\eean{\end{eqnarray*}}
\def\bfr{\begin{flushright}}\def\efr{\end{flushright}}
\def\bc{\begin{center}}\def\ec{\end{center}}
\def\bal{\begin{align}}\def\eal{\end{align}}
\def\ba#1{\begin{array}{#1}} \def\ea{\end{array}}
\def\bd{\begin{description}}\def\ed{\end{description}}
\def\bv{\begin{verbatim}}\def\ev{\end{verbatim}}
\def\nn{\nonumber}
\def\Halmos{\hfill\vrule height10pt width4pt depth2pt \par\hbox to \hsize{}}
\def\pref#1{(\ref{#1})}

\def\ins#1#2#3{\vbox to0pt{\kern-#2 \hbox{\kern#1 #3}\vss}\nointerlineskip}

\newdimen\xshift \newdimen\xwidth \newdimen\yshift
\newcount\griglia

\def\insertplot#1#2#3#4#5#6{%
\xwidth=#1pt \xshift=\hsize \advance\xshift by-\xwidth \divide\xshift by 2%
\begin{figure}[ht]
\vspace{#2pt} \hspace{\xshift}
\begin{minipage}{#1pt}
#3 \ifnum\driver=1 \griglia=#6
\ifnum\griglia=1 \openout13=griglia.ps \write13{gsave .2
setlinewidth} \write13{0 10 #1 {dup 0 moveto #2 lineto } for}
\write13{0 10 #2 {dup 0 exch moveto #1 exch lineto } for}
\write13{stroke} \write13{.5 setlinewidth} \write13{0 50 #1 {dup 0
moveto #2 lineto } for} \write13{0 50 #2 {dup 0 exch moveto #1
exch lineto } for} \write13{stroke grestore} \closeout13
\includegraphics{griglia.ps} \fi
\includegraphics{#4.ps}\fi%
\ifnum\driver=2 \fi
\end{minipage}
\caption{#5}
\end{figure}
}

\newdimen\shift \shift=-1.5truecm
\def\lb#1{%
\label{#1}\rlap{\hbox{\hskip\shift$\scriptstyle#1$}}
\else\label{#1} \fi}

\def\be{\begin{equation}}
\def\ee{\end{equation}}
\def\bea{\begin{eqnarray}}\def\eea{\end{eqnarray}}
\def\bean{\begin{eqnarray*}}\def\eean{\end{eqnarray*}}
\def\bfr{\begin{flushright}}\def\efr{\end{flushright}}
\def\bc{\begin{center}}\def\ec{\end{center}}
\def\bal{\begin{align}}\def\eal{\end{align}}
\def\ba#1{\begin{array}{#1}} \def\ea{\end{array}}
\def\bd{\begin{description}}\def\ed{\end{description}}
\def\bv{\begin{verbatim}}\def\ev{\end{verbatim}}
\def\nn{\nonumber}
\def\Halmos{\hfill\vrule height10pt width4pt depth2pt \par\hbox to \hsize{}}
\def\pref#1{(\ref{#1})}

\usepackage{amsmath}
\usepackage{amsfonts}
\usepackage{amssymb}
\usepackage{epstopdf}

\font\msytw=msbm9 scaled\magstep1 
scaled\magstep1

\let\a=\alpha \let\b=\beta  \let\g=\gamma  \let\d=\delta
\let\e=\varepsilon
\let\z=\zeta  \let\h=\eta   \let\th=\theta \let\k=\kappa \let\l=\lambda
\let\m=\mu    \let\n=\nu    \let\x=\xi     \let\p=\pi    \let\r=\rho
\let\s=\sigma \let\t=\tau    
\let\ps=\Psi   \let\o=\omega
\let\G=\Gamma \let\D=\Delta  \let\L=\Lambda

 \def\VV{{\cal V}}

\def\RR{{\cal R}}\def\LL{{\cal L}}

\def\qq{{\bf q}} \def\pp{{\bf p}}
 \def\xx{{\bf x}} \def\yy{{\bf y}} 
\def\kk{{\bf k}}

\def\nn{\nonumber}

 \def\ZZZ{\hbox{\msytw Z}}

\def\\{\hfill\break}
\def\={:=}
\let\io=\infty
\let\0=\noindent
\def\media#1{{\langle#1\rangle}}

\def\tende#1{\,\vtop{\ialign{##\crcr\rightarrowfill\crcr\noalign{\kern-1pt
    \nointerlineskip} \hskip3.pt${\scriptstyle #1}$\hskip3.pt\crcr}}\,}
\def\otto{\,{\kern-1.truept\leftarrow\kern-5.truept\to\kern-1.truept}\,}

\def\wh{\widehat}
\def\to{\rightarrow}

\def\qed{\hfill\raise1pt\hbox{\vrule height5pt width5pt depth0pt}}

\def\V#1{{\bf#1}}
\def\be{\begin{equation}}
\def\ee{\end{equation}}
\def\bp{\begin{pmatrix}}
\def\ep{\end{pmatrix}}
\def\bea{\begin{eqnarray}}
\def\eea{\end{eqnarray}}
\def\nn{\nonumber}
\def\pref#1{(\ref{#1})}

\def\lb{\label}
\def\eg{{\it e.g.}}

\def\Tr{\mathrm{Tr}}

\begin{document}

\title{Stability of Weyl semimetals with quasiperiodic disorder}
\author{Vieri Mastropietro}
\address{Dipartimento di Matematica, Universit\`a di Milano \\
  \small{Via Saldini, 50, I-20133 Milano, ITALY }}
\email{vieri.mastropietro@unimi.it}

\begin{abstract} 
Weyl semimetals are phases of matter with excitations
effectively described by massless Dirac fermions.
Their critical nature makes unclear the persistence of such phase 
in presence of disorder. We present a theorem ensuring the 
stability of the semimetallic phase in presence of weak 
quasiperiodic disorder. The proof relies
on the subtle interplay of the relativistic Quantum Field Theory description
combined with  number theoretical properties used in 
KAM theory. 
\end{abstract}

\maketitle

\section{Introduction} 

Conduction electrons in metals are well described by the
Schroedinger equation but in certain cases the interaction with the lattice produces an effective relativistic description in terms of massless Dirac particles; this happens, in particular,
in Weyl semimetals \cite{AMV}, which have been recently experimentally discovered
\cite{A11}-\cite{Bori}. This offers the possibility of observing the counterpart of high energy phenomena at a much lower energy scale, and to have materials 
with unusual physical properties.
The critical nature of excitations
has the effect that in several cases predictions are ambiguous and sensitive to approximations.
Indeed, while
there is agreement that at weak coupling many body interactions 
do not destroy the semimetalllic phase \cite{F2}-\cite{a1}, 
it is still subject of debate the effect of disorder. Field theoretical approaches find that 
a weak random disorder does not destroy the semimetallic phase
\cite{z0}, \cite{z1} while other studies
\cite{v0} based on
the inclusion of rare region effects lead to the opposite conclusion, namely  that
even an arbitrary weak random potential destabilizes the system. Numerical investigations
have been done for random 
\cite{v1a}-\cite{v1d} or quasiperiodic disorder \cite{P1},\cite{P2}, but
conclusions are subjected to finite size effects \cite{z2}. 

Rigorous results in this context are useful as can act as benchmark to check approximations or conjectures. 
In this paper we rigorously analyze Weyl semimetals on a lattice
in presence of a weak quasiperiodic disorder. Such disorder 
is the one realized in cold atoms experiments \cite{B0}, \cite{B11}; in addition quasiperiodic potential
can effectively describe coupled Dirac systems like Moire' superlattices \cite{B12}.
The effect of quasi-periodic potentials for quantum particles has been deeply studied
in one dimension; in the non interacting case 
a very detailed mathematical knowledge has been reached \cite{DS1},
\cite{DS2}, and recently great progress in understanding the effect of the interaction has been obtained  \cite{1aa}-\cite{2aa}. In contrast,
very little is known for higher dimensional Dirac systems,
with the exception of \cite{P1},\cite{P2} where numerical evidence of stability 
of the Weyl semimetallic phase was found.
The main difficulty of
quasiperiodic disorder is the presence of infinitely many processes involving a large exchange of momentum which, due to Umklapp and incommensurability of frequencies, connect fermions with momenta close to the Weyl points. Such processes are dimensionally relevant in the Renormalization Group (RG) sense and
the effect of disorder in principle increases at each RG iteration and could destroy the Weyl semimetallic phase. This phenomenon manifests in the presence in the series expansion
of small divisors which could break convergence.

A similar situation is encountered in classical mechanics and in particular in Kolmogorov-Arnold-Moser
(KAM) theory, where quasiperiodic solutions are written as Lindested series
see \eg \cite{D}. Such series are plagued by small divisors but their convergence
is ensured by subtle cancellations due to number theoretical properties of irrational numbers, see \eg \cite{GM}. In this paper we show that a similar phenomenon
allows to prove the stability of the semimetalic phase in Weyl semimetals; number theoretical
properties allow to prove that the relevant terms almost connecting Weyl points are indeed ineffective.
Physical quantities are written as convergent series so that
non-perturbative effects due to small divisors are excluded.   

The paper is organized in the following way. In \S II the model is presented, in \S III we describe the effect of Umklapp terms, in \S IV we recall number theoretical properties of irrationals and in \S V the main result is presented. Finally in \S VI the Renormalization Group analysis is presented
and \S VII is devoted to conclusions. 
\section{Weyl semimetals with quasiperiodic disorder} 
A basic model for Weyl semimetals, see \cite{AMV}, is obtained assuming a pair of orbitals
on each site of a lattice, preserving
inversion but with 
broken time reversal symmetry;  if $x=(x_1,x_2,x_3)$
are points in a cubic three-dimensional lattice $\L$, $a^\pm_{x,1}, a^\pm_{x,2}$
fermionic creation or annihilation operators, the hopping Hamiltonian is 
$H_0=$
\begin{eqnarray}&&\sum_{x\in\L}\{ \sum_{j=1}^2 (-1)^{j-1}\big[(\zeta-1)a^\dagger_{x,j}a_{x,j}+\frac12a^\dagger_{x,j}(-\Delta a)_{x,j}\big]+\\
&&\frac{it_1}2 \big[a^\dagger_{x,1}(a_{x+e_1,2}-a_{x-e_1,2})+a^\dagger_{x,2}(a_{x+e_1,1}-a_{x-e_1,1})\big]+\nn
\eea
\be
\frac{t_2}2 \big[a^\dagger_{x,1}(a_{x+e_1,2}-a_{x-e_1,2})-a^\dagger_{x,2}(a_{x+e_1,1}-a_{x-e_1,1})\big]\}\nn\ee
where in the first line $\Delta$ is the standard lattice Laplacian: $\Delta f(x)=\sum_{l=1}^3 [f(x+e_l)+f(x-e_l)-2 f(x)]$.
The Hamiltonian $H_0$ in Fourier space can be written as $H_0=\int\frac{dk}{(2\pi)^3}\,\hat a^\dagger_{k} h(k) \hat a_k$ with
\be 
h(k) =\begin{pmatrix} \alpha(k) & \beta(k) \\ \beta^*(k) & -\alpha(k)\end{pmatrix}\ee
where $k\in (0,2\pi]^3$,
$\alpha(k)=2 + \zeta-\cos k_1-\cos k_2-\cos k_3$ and $\beta(k)=t_1\sin k_1-it_2\sin k_2$. We assume that $\zeta\in [0,1)$, in which case 
$\hat h(k)$ is singular at $k=\pm p_F$, with $p_F=(0,0,\arccos\zeta)$ called Weyl point. In the vicinity of $\pm p_F$, $k=q\pm p_F$
\be
\hat H^0(q\pm p_F)=t_1 \s_1 q_1+t_2 \s_2 q_2\pm \sin p_F \s_3 q_3+
O(q^2)\label{ssss}\ee
We include now a many body interaction and quasiperiodic disorder writing
\be
H=H_0+\e\sum_{x} \phi_{x} (a^+_{x,1} a^-_{x,1}-a^+_{x,2} a^-_{x,2})
+\l\sum_{x,y} 
v(x-y) \r_{x} \r_{y}\label{h}
\ee
where $v(x-y)$ is a short range potential and 
\be
\phi_{x}=\sum_{n}\hat\phi_{n}
 e^{i 2 \p (\o_1 n_1,x_1+\o_2 n_2x_2+\o_3 n_3 x_3)}  \label{lll}
\ee
with $n\in \ZZZ^3$, $\hat\phi_n=\hat\phi_{-n}$ and  
$
|\hat\phi_{n}|\le C e^{-\x(|n_1|+|n_2|+|n_3|)}$. We assume 
the periodicity of the potential incommensurate with the lattice periodicity, by taking $\o_i$ 
{\it irrational}. The above potential includes the basic example of 
disorder like $\sum_i \cos (\o_i x_i)$. 

If $\psi^\pm_{\xx}=e^{H x_0} \psi^\pm_{x}e^{-H x_0}$, $\xx=(x_0,  x)$,
$x_0$ the imaginary time,
the 2-point function is given by
$
S(\xx,\yy)={\Tr e^{-\b H}T \psi^-_\xx\psi^+_\yy\over \Tr e^{-\b H}}$
and $\hat S(\kk)$ is the Fourier transfiorm.
In the non-interacting case $\l=\e=0$ one has $S(\xx,\yy)|_{0}=g(\xx-\yy)$
with 
\be
g(\xx)={1\over L^3 \b} \sum_\kk e^{i\kk \xx}(-i k_0 I+h(k))^{-1}
\ee
%
From \pref{ssss} we see that close to the Weyl momenta the propagator $\hat g(\qq\pm\pp_F)$ is equal to the massless Dirac propagator up to corrections. By this, one can easily
deduce the physical properties; for instance the real part 
of the zero temperature optical conductivity vanishes linearly with the frequency
$\s(\o)\sim\o $. 

In order to investigate the stability of the Weyl semimetallic phase in presence of incommensurate potential, it is convenient to write the interacting correlations 
as $S(\xx,\yy)={\partial^2 W\over\partial\phi^-_\xx\partial\phi^+_\yy}$, where 
$W(\phi)$ is Grassmann integral defined in the following way
\be
e^{W(\phi)}=\int P(d\psi) e^{V}
\ee
where $\phi$ is an external field, $\psi^\pm_{\xx,i}$ are Grassmann variables,
$P(d\psi)$ is Grassman integration with propagator
$g(\xx)$ and
\bea
&&V=\l \int d\pp \hat v(\pp) \hat\r_\pp \hat\r_{-\pp}+
\int d\xx (\psi^+_\xx\phi^-_\xx+\psi^-_\xx\phi^+_\xx)+\\
&&
\e\sum_{n,i} \hat\phi_n \int d\kk
(-1)^i \hat\psi^+_{i,\kk_1} \hat\psi_{i,\kk_2}
\d_p(\kk_1-\kk_2+\bar \o_n 2 \pi) \nn\label{bb}
\eea
where 
$\bar\o_n=(0,\o_n)$, $\o_n=(\o_1 n_1,\o_2 n_2,\o_3 n_3)$, 
$\r_\xx=\psi^+_{\xx,1}
\psi^-_{\xx,1}+\psi^+_{\xx,2}
\psi^-_{\xx,2}$, $\xx=(x_0,x)$ and $\int d\xx=
\int_{-\b/2}^{\b/2} dx_0 \sum_{\vec x}$, $\hat\r_\pp=
\int d\kk (\hat\psi^+_{\kk,1}\hat\psi^-_{\kk+\pp,1}+\hat\psi_{\kk,2}\hat\psi^+_{\kk+\pp,2}$; moreover 
\be \d_p(\xx)=\d_p(x_0) \prod_{i=1}^3 \d_p(x_i) \quad \d_p(x_i)=L\sum_n \d_{x_i, 2n \pi}\ee
Note that momentum is conserved to momenta $2\pi n$ due to the presence of the lattice.

\section{Relevant processes and Umklapp terms}

A natural way to understand the effect of the interaction and disorder is to use Renormalization Group.
The physical informations are encoded
in the marginal or relevant processes, that is the terms with vanishing or positive scaling dimension. The linear divergence at the Weyl points of the propagator 
\pref{ccc} says that the scaling dimension of the interactions with $n$ $\psi$ fields is
$
D=4-3 n/2
$, so that the only relevant terms are the bilinear ones. 
In absence of quasiperiodic potential  $\e=0$, there is only one  relevant term
corresponding to a shift in the position of the Weyl points.
The irrelevance of the quartic terms has the effect that, in the weak coupling regime, the 
semimetallic behavior persists and the only effects of the interaction are finite 
renormalization  of the velocities and wave function, see \cite{a1}.

The presence of quasi-periodic potential produces infinitely many relevant terms
quadratic in the fields, with momenta $\kk_1,\kk_2$ such that
$
k_{1,i}-k_{2,i}+2\o_i n_i \p+2  l_i \pi=0$
with $l_i, n_i$ positive or negative integers. The factor $2\o_i n_i \p$ is the momentum 
exchanged with the quasiperiodic disorder while the factor $2  l_i \pi$ is exchanged with the lattice
(Umklapp). 
Only the terms connecting fermions with momenta close to the Weyl points are really important and, due to Umklapp, this can happen also in correspondence of a non vanishing transfer of momentum
produced by the disorder. The important processes 
involve fermions with momenta close to the same Weyl  point $\s=0$ or to opposite ones
 $\s=2$ ones; if $p_F=(0,0,p_{F,3})$ this requires
\be
n_1 \o_1- l_1\sim 0\quad n_2 \o_2- l_2\sim 0\quad 
n_3 \o_3- l_3\pm \s p_{F,3}\sim 0\label{ss1}
\ee
Note the basic difference between periodic or quasi-periodic potentials.
In the first case $\o_i$ is rational $\o_i=p/q$ so that the differences in 
\pref{ss1} 
either are exactly vanishing or are  $O(1/q)$ (if $p_F\not= n \o_3/2$,): there are no processes connecting momenta arbitrarily close to the
Weyl points except the one with $n_i=0$, a process corresponding to the shift of the chemical potential.
Therefore a periodic potential is not expected to modify the physical behavior for generic values of $p_F$, at least for small $\e$.

In contrast, in the quasiperiodic case \pref{ss1} 
can be arbitrarily close to zero, for the basic properties of irrational numbers. This means that there are
infinitely many relevant processes connecting the Weyl points.
Such a feature makes the case of quasiperiodic potentials very close to the random case, where the difference of momenta of relevant terms is 
$k_1-k_2=p$ with $p$ the momentum carried by a random field $\hat\phi_p$ which can be arbitrarily small. 

\section{KAM theorem and Diophantine conditions}

In the case of random potential the issue of stability is related to the probability
that certain dangerous configurations happens. In the quasiperiodic case, the problem is deterministic and related to the irrationality properties of the frequencies. 
Therefore quantitative estimates saying how much an irrational is close to a rational one are necessary.
For instance the golden number
$\o={\sqrt{5}-1\over 2}$ verifies
$
|q \o-p|\ge {1\over (3+\sqrt{5}) } { 1\over 2\pi q }$.
If such $\o$ is the frequency of the quasiperiodic potential, this
says that, looking at
\pref{ss1}, {\it only} the processes involving a {\it large} transfer of momentum
can involve fermions {\it close} the Weyl points.
Such a property is indeed generic. There 
is a class of irrationals called
{\it Diophantine}, such that, for $q\not=0$, $p,q\in \ZZZ^2/(0,0)$
\be
|q \o-p|\ge { C_0\over 2\pi q^{\t} }\label{d1}
\ee
The irrationals not verifying \pref{d1} in the unit segment have measure $O(C_0)$; as $C_0$ can be taken arbitrarily small, the set of Diophantine numbers is full, see e.g. \cite{D}. Indeed
the set of $\o$ in the unit cube
verifying $|q \o-p|< {C_0\over ^{\t} }$ for a certain $q,p$ is 
smaller than $2 C_0/q^{\t+1}$ hence summing over $p$ (a sum bounded by $C |q|$)
and $q$ we get a set with measure bounded by $C_0 \sum_q {1\over q^\t}$ which is $O(C_0)$ for $\t>1$.  

It is therefore
not restrictive to assume the following conditions on the frequencies 
\be
|\o n|_T\ge { C_0\over |n|^{\t} }\quad 
|\o n\pm 2 p_{F,3}|_T\ge { C_0\over |n|^{\t} } \quad n\in \ZZZ/0\label{d2}
\ee
where by $|.|_T$ we mean the average on the torus, that is $|\o n|_T=\inf_{p} |\o n-p|$
; the first condition 
is \pref{d1} and the second is a requirement of irrationality for $p_{F,3}$.
As we will see, Diophantine conditions are crucial 
prove the stability of the Weyl semimetallic phase.


Another point to stress is that
in order to impose periodic boundary condition we have to choose a sequence of $\o$ rational converging to an irrational in the infinite volume limit.
In order to do that we start from 
the continued fraction representation  of a number $\o$
\be
\o=a_0+{1\over a_1+{1\over a_2+{1\over a_3+...}}}
\ee
We approximate
$\o$ by a sequence of rational numbers ({\it convergents}) 
${p_1\over q_1}=a_0+{1\over a_1}$, ${p_2\over q_2}=a_0+{1\over a_1+{1\over a_2}}$ and so on. Properties of the convergents imply
that if $\o$ verifies the Diophantine condition
then $|\pi(n{p_i\over q_i}-k)|\ge {C\over 2 |n|^\t}$ if $q_1\le n\le {q_i\over 2}$ and any $k$. Therefore we can impose periodic boundary conditions by considering a sequence of  frequencies $\o_{i}={p_i\over q_i}$ and $L_i=q_i$. 

Finally it is worth to recall that 
number theoretical conditions are unusual in condensed matter but rather common in 
other branches of physics.
For instance planets around sun neglecting the mutual  attraction have an integrable Hamilltonian dynamics which is quasiperiodic, and according to KAM theory only
quasiperiodic motions with Diophantine frequencies survive in presence of perturbation breaking integrability \cite{D}.
Indeed quasi periodic solutions are written as series in the perturbation, called Lindstedt series,
whose convergence follows by subtle cancellations due to Diophantine conditions, see \eg \cite{GM}.

\section{Main result} 

As the interaction in general moves the location of the Weyl mometum, we write
$\x=\cos p_F+\n$ in \pref{h} and we choose $\n$ so that $p_F$ is the just the interacting Weyl momentum.  
\vskip.2cm
{\bf Theorem.}{\it
For $\l,\e$ small enough and assuming that the frequencies $\o_i$ in 
\pref{lll} verify
\pref{d2}, there exists $\n$ such the 2-point function $\hat S(\kk)$ behaves as, if $\pp_F=(0,p_F)$, 
\be
S(\qq\pm \pp_F)= {1\over Z }\begin{pmatrix} -i q_0\pm v_3 q_3 & 
v_1 q_1-i v_2 q_2 \\ v_1  q_1+iv_2 q_2 & -iq_0\mp v_3 q_3 
\end{pmatrix}^{-1}(1+O(\qq))\label{ccc}\nn
\ee
with $Z=1+O(\l,\e)$, $v_1=t_1+O(\l,\e)$, $v_2=t_2+O(\l,\e)$, $v_3=\sin p_F+O(\l,\e)$
}
\vskip.1cm
This result proves the stability of the Weyl semimetallic phase, as 
quasiperiodic disorder does not modify qualitatively
the 2-point function but produces only
a finite renormalization of the parameters; no phase transition is present at small disorder.
As a consequence the real part of optical conductivity 
vanishes as $O(\o)$ as in the non interacting case. Even if 
there are infinitely many relevant terms due to quasiperiodic disorder, they do not modify the physical behavior. 
The result is in agreement with the numerical evidence in \cite{P1},\cite{P2}.

\section{Renormalization Group} 

In order to prove \pref{ccc} we need to evaluate the generating
function $\int P(d\psi)e^\VV$ with $\VV=V+\n \int \hat\psi^+\s_3\hat\psi^-
$ with $V$ given by 
\pref{bb} and propagator given by $g(\xx)$.
We introduce two smooth cut-off functions $\chi_\pm(\kk)$ non vanishing in a region $|\kk\mp \pp_F|\le \g$ and
non-overlapping, $\g$ a suitable constant:
we define $\hat g_\r^{(\le 0)}(\kk)=\chi_\r(\kk)\hat g(\kk)$
and 
\be
g(\xx)= g^{(1)}(\xx)+\sum_{\r=\pm} g^{(\le 0)}_\r(\xx)
\ee
with $\hat g^{(1)}(\kk)=(1-\sum_\r\chi_\r)\hat g(\kk)$; this induces the Grassmann variable decomposition $\psi_\xx=\psi^{(1)}_\xx+\sum_{\r=\pm} \psi^{(\le 0)}_\r$
with propagators given by  $g^{(1)}(\xx)$ and $g^{(\le 0)}_\r(\xx)$
respectively. Note that $\psi^{(1)}$ correspond to fermions with momenta far from the Weyl points, while  $\psi^{(\le 0)}_\pm$ with momenta around $\pm \pp_F$. 

We can further decompose $\hat g_\r(\kk)=\sum_{h=-\io}^0 
\hat g^{(h)}_{\r}( \kk)$, $\r=\pm$ with the cut-of function
$\chi_\r$ replaced by $f_h$ with support in $ \g^{h-1} \le |\kk-\r\pp_F|\le  \g^{h+1}$.
After the integration of $\psi^{(1)}, \psi^{(0)},..,\psi^{(h+1)}$ the generating function has the form
\be
e^{W(\phi,J)}=\int P(d\psi^{(\le h)})e^{\VV^{(h)}(\psi^{(\le h)},\phi)}\label{hh}
\ee
where $P(d\psi^{(\le h)})$ has propagator 
\be
\hat g^{(h)}_\pm(\qq)={1\over Z _h} f_h(\qq)\begin{pmatrix} -i q_0\pm v_{3,h} q_3 & 
 v_{1,h} q_1-i v_2 q_2 \\  v_{1,h}  q_1+iv_2 q_2 & -iq_0\mp  v_{3,h} q_3 
\end{pmatrix}^{-1}\label{ccc1}\nn
\ee
and $\VV^{(h)}(\psi,0)=$
\be \sum_{m, n,\underline\r} \int d\qq_1...d\qq_m  W_{ n,m}^{(h)}(\underline \qq)\psi^{\e_1(\le h)}_{\r_1,\qq_1}...\psi^{\e_m(\le h)}_{\r_m,\qq_m}\d_{n,m}(\underline\qq)
\label{ep}\ee
where $\d_{n,m}(\underline\qq)$ is $L\b$ times a periodic Kronecker delta non vanishing for $\sum_{i=1}^{m}\e_i  q_{0,i}=0$ and, $p_F=(0,0,p_{F,3})$
\be
\sum_{i=1}^{m}\e_i q_i=-\sum_{i=1}^{m}\e_i \r_i  p_F
+2 \pi  \o_{n}+2 l \pi \label{hh1}
\ee
with $l=(l_1,l_2,l_3)$ and $\o_n=(\o_1 n_1, \o_2 n_2, \o_3 n_3)$.
$\VV^{(h)}(\psi,\phi)$ has a similar expression with some $\psi$ field replaced by an external field. The stability of the semimetallic phase relies in the fact
that the sequence of effective potentials $\VV^h$ remains small for any RG iteration.
There are however relevant terms, that is terms that could increase linearly according to dimensional analysis; they are the infinitely many terms, depending on $n$, 
with $m=2$ in \pref{ep}. One neeeds to show that, despite the linear divergence suggested by scaling, such terms indeed
remain small due to cancellations relying on number theoretical properties.

Note first that one can distinguish between the terms with $m=2$ such that the l.h.s.
of \pref{hh1} is vanishing, which we call resonant, from the other, which we call non-resonant. The resonant terms with $m=2$ are possible only for $n_i=0$, $l=0$ and $\r_1=\r_2$;
the case $\r_1=-\r_2$ would be possible if $p_{F,3}=n \o_3/2$, a case excluded by 
the assumption \pref{d2}. We define a localization operator
\be
\LL W_{ 0,2}^{(h)}(\qq)=
W_{ 0,2}^{(h)}(0)+\qq \partial W_{ 0,2}^{(h)}(0)
\ee 
and $\LL W_{ n,m}^{(h)}(\qq)=0$ otherwise.
Note that the graphs contributing
to the nondiagonal part contain an odd number of non diagonal propagators hence are vanishing; moreover $W_{ 0,2;11}^{(h)}(0)=-W_{ 0,2;22}^{(h)}(0)$.
In addition the derivative with respect to $0,3$ of graphs contributing to the non diagonal part 
is zero, as they contain an odd number of non diagonal propagators, and the derivative with respect to $1,2$ of graphs contributing to the diagonal part 
is zero, as it contain an even number of non diagonal propagators. 

We can write
$\VV^h=\LL \VV^h+\RR \VV^h$ with $\RR=1-\LL$ and rewrite \pref{hh} as
\be
\int P(d\psi^{(\le h)})e^{\g^h \n_h F^{(h)}+\RR\VV^{(h)}(\psi^{(\le h)},\phi)}\label{hh}
\ee
with $F^{(h)}=\int d\xx (\psi^+_{\xx,1}\psi^-_{\xx,1}- \psi^+_{\xx,2}\psi^-_{\xx,2} )$.
One can write $P(d\psi^{(\le h)})=P(d\psi^{(\le h-1)})P(d\psi^{(h)})$ and integrate $\psi^{(h)}$
obtaining an expression similar to \pref{ccc1} with $h-1$ replacing $h$, and the procedure can be iterated.

The kernels $W^{(h)}_{n,m}$ can be written as sum of Feynman diagrams composed by vertices connected by lines,
such that to each line is associated a scale label $h$ and it corresponds to a propagator $g^{(h)}$.

\insertplot{230}{100}
{\ins{24pt}{67pt}{$v_1$}
\ins{80pt}{67pt}{$v_2$}
\ins{50pt}{77pt}{$v_3$}
\ins{110pt}{67pt}{$v_5$}
\ins{100pt}{77pt}{$v_4$}
}
{figjsp44ac}
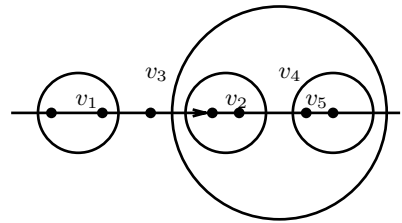
{\label{h2} An example of graph of order $\e^7$ with the associated clusters ( the circles)
contributing to
$\hat W^{(h)}_{0,2}(\kk)$; the value is $\hat g^{h_{v_1}}(\kk_1)\hat g^{h_{v_3}}(\kk_2)
\hat g^{h_{v_2}}(\kk_3)\hat g^{h_{v_4}}(\kk_4)\hat g^{h_{v_5}}(\kk_5)$
with $k_i=k+2\pi \o_{n_i}$ and $h_{v_3}<h_{v_1}$, $h_{v_3}<h_{v_4}$, $h_{v_4}<h_{v_2}$, $h_{v_4}<h_{v_5}$.
} {0}

We can therefore consider a cluster $v$, that is
the maximally connected subset of lines corresponding 
to propagators 
with scale $h\ge h_v$ with at least a scale $h_v$; 
$n^e_v$ is the number of lines external to the cluster $v$, by definition 
with scale smaller then $h_v$ (for more details see e.g. \cite{M1}).
The structure of clusters induce a natural notion of subgraphs which avoid the well known problem of overlapping divergences.
The clusters with $n^e_v=2$
are such that the difference of momenta $k$ is $2\pi (N_1 \o_1, N_2\o_2,N_3 \o_3)$.
We call $v'$ the minimal cluster containing $v$ so that 
$h_v-h_{v'}>0$,
and $S_v$ is the number of clusters or vertices contained in $v$ and not in any smaller cluster. 

Using that the propagator $g^{(h)}(\xx)$ is bounded by $\g^{3 h}$ and the integral
of the propagator  over coordinates by $\g^{-h}$, a graph is bounded by $C^s (\max(\e,\l,\n_h)^s$ times
\be
\prod_v \g^{-4 h_v(S_v-1)} \prod_v \g^{ 3 h_v n_v} \prod_v \g^{z_v (h_{v'}-h_v)}
\ee
where $n_v$ is the number of propagators in the cluster $v$ but not in any smaller one, and 
$z_v=2$ is $v$ is a resonant cluster with $n^e_v=2$ and zero otherwise; 
the last term in the above expression is produced by the renormalization $\RR$.

By using the relations \bea
&&\sum_v (h_v-h)(S_v-1)=\sum_v (h_v-h_{v'})(m^4_v+m^2_v-1)\nn\\
&&\sum_v (h_v-h) n_v=\sum_v (h_v-h_{v'})(2 m^4_v+m^2_v-n^e_v/2 )\nn\eea
one gets 
\be \g^{ D h}
\prod_v \g^{(h_v-h_{v'})(D_v-z_v)
}\prod_v \g^{  2 h_v \bar m^4_v} \prod_v \g^{-h_v \bar m^2_v}
[\prod_{i} e^{-\x |\vec n_i|} ]\label{paz}\ee
where $\bar m_v^4$  is the number of vertices $\l$ contained in $v$ and not in any smaller cluster, $\bar m_v^2$
is the number of vertices $\e$ contained in $v$ and not in any smaller cluster, $s$ is the order,
$D_v=4-3 n^e_v/2$; the last term is due to the decay factors $\hat\phi_n$.
Note that the same bound is valid for the sum over all Feynman graphs form the cancellations due to Pauli principle; no combinatorial problems arise for the number of graphs, see \cite{M1} 

One needs to sum over all the choices of scales $\{ h\}$.
By looking to \pref{paz} we see indeed that if for all $v$ one has $D_v-z_v<0$
then one can sum over all the choices of scales, that is 
$\sum_{ \{ h\}}
\prod_v \g^{(h_v-h_{v'})(D_v-z_v)}$ is bounded by $C^s$ (remember that $h_v-h_{v'}>0$).
There are however clusters with $D_v-z_v=0$, actually the non resonant clusters with $n^e_v=2$, and this produces a divergent bound
$|h|^s$ . Such divergence may suggests that the Weyl semimetallic behavior is unstable.

We need however to take into account the 
number theoretical properties of the frequencies.
Let us consider a subgraph with two external lines (see e.g. Fig. 2), associated to propagators
with momenta $\kk_1,\kk_2$. 
If $q_1, q_2$ are the momenta measured from Weyl points,
$k=q\pm p_F$, external to a cluster $v$,
one has
$|q|\le \g^{h_{v'}}$ for the compact support properties of the propagator. 
Moreover $k_1-k_2=2\pi (N_1\o_1, N_2\o_2, N_3 \o_3)$
so that, if $|q|_T=\sqrt{|q_1|_T^2+|q_2|_T^2+|q_3|_T^2}$
\be
2\g^{h_{v'} } \ge |q_1|_T+|q_2|_T\ge |q_1-q_2|_T\ee
where we have used the triangular inequality on the torus.
Now we use the Diophantine property \pref{d2}, $\e=0,\pm$
\be
2\g^{h_{v'} }\ge \sqrt{
|\o_1 N_1|_T^2+|\o_2 N_2|_T^2+|\o_2 N_3+\e 2 p_{F,3} |_T^2}\ge { 3 C_0\over  \bar N^\t }\label{lau}
\ee
so that, if $\bar N=\max (N_1,N_2,N_3)$ then 
\be
\bar N\ge C \g^{-h_{v'} /\t}
\ee
This inequality says that if the momenta external to
a non resonant cluster are very small, than the momentum transferred is very large. 
On the other hand by conservation of momentum $N_v=\sum_i n_i$ where $n_i$ is the momentum associated with each $\e$ vertex in the cluster, and \be
\prod_i e^{-\x|n_i|}
\le e^{-\x \bar N}\le e^{-C \g^{-h_{v'} /\t}}\ee
as $\sum_i |n_i|\ge |\sum_i n_i|\ge \bar N$. This relation implies a dramatic improvement with respect to the dimensional bound.
 
\insertplot{320}{90}
{\ins{2pt}{70pt}{$\kk_1$}
\ins{52pt}{70pt}{$\kk_2$}
\ins{100pt}{70pt}{$\kk_3$}
\ins{24pt}{77pt}{$>$}
\ins{24pt}{47pt}{$<$}
}
{figjsp44ad}
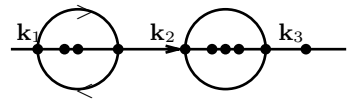
{\label{h2} A contribution order $\l^4 \e^6$ to $\hat W^{(h)}_{0,2}(\kk)$; if $k_2-k_3=2\pi\o_n$ and
$|q_2|\sim \g^{-h_{v'}}$, $|q_3|\sim \g^{-h_{v'}}$ then $max_i |n_i|\ge \g^{-h/\t}$.
} {0}

There is however in general a sequence of clusters enclosed one in the other and $\e$ vertices are generally internal to several clusters. In order to get a decay factor for each cluster we can write
\be
e^{-\x |n|/2}=\prod_{h=-\io}^{-1} e^{-\x 2^h |n|/2}
\ee
We can therefore associate
to each relevant non-resonant cluster $v$ a factor $e^{-\x \bar N_v 2^{h_{v'} } }$, so that, see Fig. 3
\be
e^{-\x \bar N_v 2^{h_{v'} } }\le e^{-\x   2^{h_{v'}} \g^{-h_{v'}/\t}  }\ee
If we choose  $\g^{1/\t}=4$ then  $e^{-\x   2^{-h_{v'}}}\le (N e/\x)^N 2^{N 2h_{v'}}$, by using  $e^{-\a x} x^N\le ({N e\over \a})^N$.
\insertplot{230}{110}
{\ins{40pt}{40pt}{$v_1$}
\ins{60pt}{40pt}{$v_2$}
\ins{45pt}{60pt}{$v_3$}
\ins{210pt}{40pt}{$+...$}
}
{figjsp44ab}
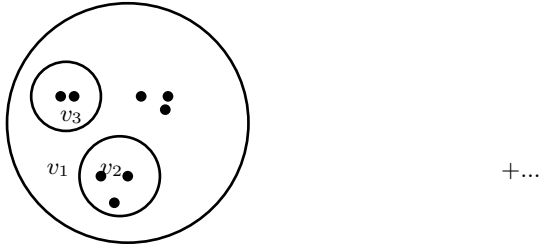
{\label{h2} Three clusters $v_1,v_2,v_3$.
If the points are associated to $\phi_{n_i}$, assume $1,2,3$ in $v_2$, $3,4$ in $v_3$ and $4,5,6$ in $v_1$. Hence
$\prod_{i=1}^8 e^{- |n_i|}$ is bounded by $e^{-(|n_1|+|n_2|+|n_2|) 2^{h_{v_2}}}
e^{-(|n_3|+|n_4|) 2^{h_{v_3}}} e^{-(|n_1|+...|n_8|) 2^{h_{v_1}}}$
} {0}
Therefore, choosing $N$ so that  $2^N=\g$ ($N=2\t$) 
\be
[\prod_{i} e^{-\x |\vec n_i|/2} ]\le C^s \prod_{v} \g^{h_{v} 2 S_v^{NR}}
\ee
where $S_v^{NR}$ is the number of non resonant clusters or vertices in $v$ and
not in any smaller cluster and $m$ is the order. Using that
\be
\prod^*_v \g^{-2(h_{v'}-h_v)
}\prod_v \g^{-h_v 2 \bar m^2_v}
\le \prod_v \g^{-h_{v} 2 S_v^{NR}}\ee
where the first product
 is over the non resonant relevant $v$
we get that \pref{paz} is replaced by
\be \g^{ D h}
\prod_v \g^{(h_v-h_{v'})(D_v-\bar z_v)
}\prod_v \g^{h_v \bar m^4_v} 
[\prod_{i} e^{-\x |\vec n_i|/2} ]\label{paz}\ee
with $\bar z_v=2$ for $n^e_v=2$ and zero otherwise.

As  $D_v-\bar z_v>1$
we can sum over all the scale choices getting a bound $O(C^s max(\l,\e,\n_h)^s)$ from which convergence of the series expansions follows, provided that $\n$ is chosen so that $\n_h$ vanishes as $h\to-\io$. 

Finally we note that the velocities verify a recursive relation
$v_{h-1}=v_h+\b_{v}^h$ and the wave function renormalization verifies $Z_{h-1}=Z_h+\b_z^h$;
note that the Feynman graphs contributing to $\b_h$ have at least a $\l$ vertex so that 
$\b^h=O(\l\g^h)$ and $v_{-\io}=v_0+O(\l)$, $Z_{-\io}=1+O(\l)$.
The correlations are therefore close to the non-interacting ones up to finite renormalizations.

\section{Conclusion} 

We have rigorously established the stability of the Weyl semimetallic
phase in presence of weak interaction and quasiperiodic disorder. Even if the infinitely many relevant terms produced by the disorder could possibly destabilize the semimetallic phase, this is avoided 
by subtle cancellations due to number theoretical properties. The physical properties appear to be determined by the interplay of relativistic Quantum Field Theory with
classical mechanics and KAM theory. There are no phase transition for
weak quasiperiodic disorder, where rare region effects are absent. If a similar rigorous RG analysis can be performed in the case of random disorder is a very interesting open question.

{\bf Acknowledgements.} This work has been supported by MIUR, PRIN 
2017 project MaQuMA. PRIN201719VMAST01.

\bibliographystyle{amsalpha}

\end{document}